\documentclass[showpacs,pre,12pt]{revtex4}

\usepackage{latexsym,amssymb,amsmath}
\usepackage{graphics}
\usepackage{graphicx}
\usepackage{epsfig}

\begin{document}

\title{Condensation Transitions in Two Species Zero-Range Process}

\author{T. Hanney and M. R. Evans}

\affiliation{School of Physics, University of Edinburgh, Mayfield
Road, Edinburgh, EH9 3JZ, UK} 

\begin{abstract}
We study condensation transitions in the steady state of a zero-range
process with two species of particles. The steady state is exactly
soluble --- it is given by a factorised form provided the dynamics
satisfy certain constraints --- and we exploit this to derive the
phase diagram for a quite general choice of dynamics. This phase
diagram contains a variety of new mechanisms of condensate formation,
and a novel phase in which the condensate of one of the particle
species is sustained by a `weak' condensate of particles of the other
species. We also demonstrate how a single particle of one of the
species (which plays the role of a defect particle) can induce
Bose-Einstein condensation above a critical density of particles of
the other species.  
\end{abstract}


\maketitle


\section{Introduction}

Condensation phenomena are observed in a variety of contexts.
For instance, microscopic dynamics including particle diffusion,
aggregation to form particle clusters, and fragmentation of these
clusters, can be used to model a number of
physical systems \cite{MKB,KR}. In one dimension, such models have been
analysed within mean field theory \cite{MKB} or using a scaling approach
\cite{KR} to infer the existence of transitions between a fluid phase a
condensate phase. Another context is the modelling of granular and
traffic flow \cite{E96}. One such model is the `Bus Route model' \cite{OEC},
where there is a crossover between a regime in which the average
velocity of buses is determined by the velocity of the slowest bus,
and, above a critical density of buses, a regime in which the average
bus velocity is limited instead by the high density of traffic. This
crossover can be understood in terms of a condensation process. 
Further, several models have been shown to undergo phase separation in
one dimension \cite{AHR,M}. This phase separation can also be related to a
condensation mechanism. In particular, a general criterion has been proposed,
predicting the existence of phase separation in 1-$d$ driven systems,
which appears to be widely applicable \cite{KLMST}. This applicability
follows from the robust nature of the physical mechanism underlying the phase
separation. This mechanism, which is also the generic mechanism for
the aforementioned condensation phenomena, is understood in terms of
condensation transitions in the zero-range process \cite{E00}.

The zero-range process is a system of many interacting particles which
move on a lattice --- particles hop to adjacent lattice sites with
hop rates determined by the number of particles present at the
departure site. It provides insight into the behaviour of more
complicated models because it is exactly soluble: the steady state
assumes a simple, factorised form. Thus the condensation transitions,
whereby a finite fraction of particles occupy a single site,
are amenable to exact analysis. 
Condensation transitions in the single species zero-range process
proceed through one of two mechanisms: (i) if the particle hop rates
are site dependent, then above a critical density a
condensate forms at the site where the hop rate is slowest --- this
mechanism is closely related to Bose condensation, (ii) if
the particle hop rates depend on the number of particles present at
the departure site, then above a critical density, and provided the
asymptotic dependence decays to a constant value sufficiently quickly,
a condensate forms at a site located at random --- thus the
transition is accompanied by a spontaneously broken symmetry. The
former mechanism here demonstrates how disorder can induce
condensation, analysis of the latter leads to an understanding of the
generic mechanism of condensate formation applicable to the examples
described above.  

A question which naturally arises then is what other generic mechanisms of
condensate formation exist. To this end, we consider the
generalisation of the zero-range process to two species of
particles. We investigate how the interaction of the two species
allows new mechanisms of condensate formation.
  
The zero-range process with two species of particles was introduced in
\cite{EH03, GS03} where the steady state was obtained exactly and shown to
be given by a simple factorised form provided the dynamics satisfy certain
constraints. In \cite{EH03}, this was used to demonstrate a new
mechanism of condensation transition for a specific choice of
dynamics, and in \cite{GS03} the hydrodynamics were derived. Here, we
show how the two species model may undergo a wide variety of
condensation transitions, and we derive the phase diagram for the model
for a quite general choice of dynamics. Also, we show that the steady
state of the model can be mapped on to the steady state of
the AHR model \cite{AHR} --- a model which undergoes a transition between a
disordered (fluid) phase and a phase separated (condensate) phase. 
Thus, as in the single species model, the two species zero-range process
exhibits transitions of a robust nature which can provide insight
into condensation mechanisms in more complicated models.
In particular, one perspective of the two species model is to consider
one species as providing an evolving landscape upon which the other species
in turn evolves. This evolution is coupled, thus condensation transitions are
induced by the evolving disordered background. Again, such
interplay arises in a variety of physical settings
\cite{LR,DK,DB}. Our aim is to explore how this interplay can lead to
novel condensation transitions in the two species zero-range process.

We begin by reviewing in Section 2 the key equations of the steady state solution, which form the basis of the subsequent analysis. In Section \ref{DP},
we show how a defect particle (i.e. a single particle of one of
the species) can induce a condensation transition in the particles of
the other species. We consider in Section \ref{GC} the case where the
hop rates of one of the particle species depend only on the number of
particles of the other species at the departure site. We show how to
derive the phase diagram for this case and find that three distinct
condensate phases can arise; numerical iterations of an
exact recursion relation for the partition function yield results
consistent with the predicted phases. In Section \ref{AHR} we present
a mapping between the steady state of the two species zero range
process and the AHR model. We conclude in Section \ref{C}.


\section{Steady State}
\label{SS}

We define the two species zero-range process on a lattice containing
$L$ sites and with periodic boundary conditions. On this lattice,
there are $N$ particles of species $A$ and $M$ particles of species
$B$. Particles of both species hop to the nearest neighbour site to the right,
species $A$ with rate $u(n_l, m_l)$ and species $B$ with rate $v(n_l,
m_l)$, where site $l$ is the departure site and contains $n_l$
particles of species $A$ and $m_l$ particles of species $B$ . 

Since the steady state has already been derived in detail elsewhere
\cite{EH03, GS03}, we quote only the key results here. We define
$P(\{n_l\};\{m_l\})$ to be the probability of finding the system in
the configuration $(\{n_l\};\{m_l\})$, where $\{n_l\}=n_1,\ldots,n_L$ and
$\{m_l\}=m_1,\ldots,m_L$. This is given by a factorised form 
\begin{equation} \label{PNM}
P(\{n_l\};\{m_l\}) = Z_{L,N,M}^{-1} \prod_{l=1}^{L} f(n_l,m_l) \;,
\end{equation}
where $Z_{L,N,M}$ is a normalisation. The steady state (\ref{PNM})
satisfies the steady state master equation if the factors $f(n_l,m_l)$ satisfy
\begin{equation} \label{AB}
\frac{u(n_l,m_l) f(n_l,m_l)}{f(n_l\!-\!1,m_l)} = 1 \qquad {\rm and}
\qquad \frac{v(n_l,m_l) f(n_l,m_l)}{f(n_l,m_l\!-\!1)} = 1 \;.
\end{equation}
The solution to these equations is 
\begin{equation} \label{FNM}
f(n_l,m_l) = \prod_{i=1}^{n_l} \left[ u(i,m_l) \right]^{-1}
\prod_{j=1}^{m_l} \left[ v(0,j) \right]^{-1} \;,
\end{equation}
provided the hop rates satisfy the constraint
\begin{equation} \label{CE}
\frac{u(n_l,m_l)}{u(n_l,m_l\!-\!1)} = \frac{v(n_l,m_l)}{v(n_l\!-\!1,m_l)} \;,
\end{equation}
for $n_l,m_l \neq 0$ --- the choices of $u(n_l,0)$ and $v(0,m_l)$ remain
unconstrained. We emphasise that instead of specifying the hop
rates directly, we have the freedom to choose any desired form for
$f(n,m)$ and that we can then infer the hop rates from (\ref{AB}). 

The normalisation $Z_{L,N,M}$, defined in (\ref{PNM}), plays a role
analogous to the canonical partition function of equilibrium
statistical mechanics, and is given by
\begin{equation} \label{ZLNM}
Z_{L,N,M} = \sum_{\left\{ n_l \right\} , \left\{ m_l \right\}}
\delta( \sum_{l=1}^L n_l - N) \delta( \sum_{l=1}^L m_l - M)
\prod_{l=1}^{L} f(n_l,m_l) \;,
\end{equation}
where the delta-functions ensure that the system contains the correct
numbers of particles of each species. By writing the delta-functions
in an integral representation, $Z_{L,N,M}$ becomes
\begin{equation} \label{NORM}
Z_{L,N,M} = \oint \frac{dz}{2 \pi i} \oint \frac{dy}{2 \pi i}
\frac{[ F(z,y) ]^L}{z^{N+1}y^{M+1}} \;,
\end{equation}
where the generating function, $F(z,y)$, has been defined as
\begin{equation} \label{FZY}
F(z,y) = \sum_{n=0}^{\infty} \sum_{m=0}^{\infty} z^n y^m f(n,m)\;.
\end{equation}
We consider in Section \ref{GC} the limit $L,N,M \to \infty$, where
$\rho_A = N/L$ and $\rho_B = M/L$ --- the particle densities of
species $A$ and $B$ respectively --- are held fixed. In this limit, we
assume that the integral in equation (\ref{NORM}) is dominated by the
saddle point. The equations for the saddle point are
\begin{equation} \label{PD}
\rho_A = z \frac{\partial}{\partial z} {\rm ln} F(z,y) \;, \qquad
\rho_B = y \frac{\partial}{\partial y} {\rm ln} F(z,y) \;.
\end{equation}
Assuming the saddle point is valid,
equations (\ref{PD}) determine 
$\rho_A$ and $\rho_B$ in terms of $z$ and $y$
and this amounts to working in a grand canonical ensemble.
We note that for the saddle
point to be valid, $z$ and $y$ cannot exceed the radii of
convergence of $F(z,y)$, since we must be able to perform the sum
(\ref{FZY}) in the first place. Further, since all derivatives of
$F(z,y)$ are positive, the saddle point, if valid, must be unique. In
Section \ref{GC} 
we will find that it is not always possible to solve the
saddle point equations for all values of $\rho_A$ and $\rho_B$ in the
allowed ranges of $z$ and $y$. This phenomenon corresponds to a
condensation transition.


\section{Defect particle}
\label{DP}

In this section, we consider how condensation may arise when there is
only a single particle of species $B$.
The hop rates of the $A$ particles are chosen to be
\begin{equation} \label{DPU}  
u(n,0)=1 \qquad {\rm and} \qquad u(n,1)=p\;, 
\end{equation}
where $p<1$, such that the $A$ particles hop more slowly when the $B$
particle is present (but they hop independently of $n$). Thus we view
the $B$ particle as a defect particle. The constraint on the hop rates
(\ref{CE}) then requires that the hop rate of the $B$ particle is
\begin{equation} \label{DPV}
v(n,0)=0 \qquad {\rm and} \qquad v(n,1)=p^n\;.
\end{equation}
Substituting these dynamics into (\ref{FNM}), one finds that
$f(n,0)=1$ and $f(n,1)=p^{-n}$. 
For this model, we can evaluate (\ref{ZLNM}) as a finite sum
(i.e. we can work directly in the canonical ensemble).
Taking the $B$ particle to be at site $k$, and summing over the $L$
possibilities for $k$,
yields
\begin{eqnarray} 
Z_{L,N,M} &=& L \sum_{\left\{ n_l \right\}}
\delta( \sum_{l=1}^L n_l{-}N) p^{-n_k} \\
&=& L \sum_{n_k =0}^{N} \binom{L{-}2{+}N{-}n_k}{L{-}2} p^{-n_k} \;.\label{PF}
\end{eqnarray}
This
is the same normalisation (up to an overall factor of $L$) as that
derived for the single species model with heterogeneous hop rates,
in the case where particles 
hop from all sites with rate 1 except for a single defect site from
which particles hop with rate $p$ \cite{E00}. Thus, using the results
of \cite{E96}, one can identify
two regimes: a low density phase, when $\rho_A < p/(1-p)$, and the system
is in a fluid phase; and a high density phase, when $\rho_A >
p/(1-p)$, and a
Bose condensate forms at the site containing the defect particle. 
These regimes can be computed exactly by considering the normalisation
(\ref{PF}), and seeing that the sum may be dominated either by $n_k
\sim \mathcal{O}(1)$, in which case $\rho_A < p/(1-p)$, or by $n_k
\sim \mathcal{O}(L)$, in which case $\rho_A > p/(1-p)$ \cite{E00}.
The condensate then contains a finite
fraction of all the particles in the system --- the remaining
particles form a power law distributed background. It is a Bose
condensate in the sense that the particles condense onto the site
containing the defect particle (in the same way that in Bose
condensation, the particles condense into the state of lowest energy:
the equivalence is observed by identifying the site containing the
defect particle in the zero-range process with the state of lowest
energy in the Bose gas).
 
A condensation transition of this kind persists as long as we have a
finite number of (indistinguishable) defect particles: let's say there
are $M$ defect particles where in the limit $L\to \infty$, we keep $M$
fixed. Also, the hop rates for the $A$ particles are $u(n,m) = p_m$
where $m=1,\ldots,M$. If the smallest of these hop rates is $p_i$,
then above a critical density of $A$ particles, all the sites containing
$i$ particles of species $B$ contain a finite fraction of all the
particles of species $A$. But because there can only be a finite
number of such sites, each of these sites must contain an infinite
number of particles of species $A$ --- the condensate is distributed
equally among the sites containing $i$ particles of species $B$. 

The analysis of this section leads us to view the defect particle(s) 
as a disordered background upon which $A$ particles evolve. The special
feature of the two species model is that this background may also
evolve with prescribed dynamics. This is the case we consider in
the following section, when the number of $B$ particles is extensive.


\section{Finite densities of both species}
\label{GC}

With the perspective of particle dynamics on an evolving disordered
background, a case of particular interest in the two species
zero-range process is when the evolution of one species,
the $B$ particles say, depends only
on the number of particles of the other species at a site. Therefore
we take $v(n,m) = 1+r(n)$ for $m>0$, where $r(n)$ is a general function of
$n$. Then from (\ref{AB}) we deduce that $f(n,m)$ is given by  
\begin{equation} \label{GEN}
f(n,m) = [1+r(n)]^{-m} s(n)\;,
\end{equation}
where $s(n)$ is another general function of $n$. We then use
(\ref{AB}) to infer the rates $u(n,m)$:
\begin{equation}
u(n,m) = \left(\frac{1+r(n-1)}{1+r(n)}\right)^{-m}
\frac{s(n-1)}{s(n)}\;.
\end{equation}
We assume in the following that $r(n)$ is a monotonically decreasing
function of $n$, and that in the limit $n \to \infty$, $r(n) \to 0^+$. 
Inserting the form (\ref{GEN}) into (\ref{FZY}), and performing the sum over
$m$, yields 
\begin{eqnarray}
\label{fzy}
F(z,y) = \sum_{n=0}^{\infty} s(n) z^n \frac{1+r(n)}{1+r(n)-y}\;,\\
\label{zdzfzy}
z \frac{\partial}{\partial z} F(z,y) = \sum_{n=0}^{\infty} n s(n) z^n
\frac{1+r(n)}{1+r(n)-y}\;, \\
\label{ydyfzy}
y \frac{\partial}{\partial y} F(z,y) = \sum_{n=0}^{\infty} s(n) z^n
\frac{y [1+r(n)]}{[1+r(n)-y]^2}\;.
\end{eqnarray}
These equations determine $z$ and $y$, given the densities $\rho_A$
and $\rho_B$, via (\ref{PD}). 
The radius of convergence of the sum over $m$ is $y=1$
and we take, without loss of generality,
the radius convergence of the sum over $n$ to be $z =1$.

To analyse the possible transitions, we need to elucidate the behaviour of
$\rho_A$ and $\rho_B$ when considered as a function of $z$ and
$y$. This will enable us to draw 
graphs of the dependences of $\rho_A$ and $\rho_B$ on
$y$, for fixed values of $z$, from which we can determine the
densities for which the saddle point approximation remains valid. In
particular, we wish to consider how $\rho_A$ and $\rho_B$ change as
$z$ and $y$ approach their radii of convergence --- if $\rho_A$ or
$\rho_B$ tends towards a finite value, then condensation ensues.
To this end, we make the following observations: 
\begin{enumerate}
\item For fixed $z$, $\rho_A$ and $\rho_B$ are monotonically
increasing functions of $y$. 
\item For $z\to 0$, $\rho_A \to 0$.
\item For $y\to 0$, $\rho_B \to 0$.
\item For $z<1$, $\rho_B$ is finite for all $y$ (including
$y=1$).
\end{enumerate}
We supplement these observations with the following three conditions on
$r(n)$ and $s(n)$, which determine whether $\rho_A$ and $\rho_B$
converge to finite or infinite values when $z$ and $y$ approach their
radii of convergence. 
For $z \to 1$ and $y<1$, if, as $n \to \infty$, 
\begin{equation} \label{C1}
n s(n) \to 0  \quad \textrm{faster than} \quad 1/n\;,
\end{equation} 
then $\rho_A \to {\rm finite}$.
For $z \to 1$ and $y \to 1$, if, as $n \to \infty$, 
\begin{equation} \label{C2}
\frac{n s(n)}{r(n)} \to 0 \quad \textrm{faster than} \quad 1/n\;,
\end{equation}
then $\rho_A \to {\rm finite}$. 
For $z\to 1$ and $y \to 1$, if, as $n \to \infty$, 
\begin{equation}\label{C3} 
\frac{s(n)}{r(n)^2} \to 0 \quad \textrm{faster than} \quad 1/n\;,
\end{equation}
then $\rho_B \to {\rm finite}$. 
These observations and conditions enumerate all possible ways that $z$
and $y$ approach their radii of convergence, and therefore all the
possible circumstances in which condensation can occur in our
model. The phase behaviour depends on which of the conditions
(\ref{C1}) to (\ref{C3}) are met and which are not.  

There are several possibilities, which we illustrate for a particular
choice of $r(n)$ and $s(n)$, namely, for large $n$,
\begin{equation} \label{FORMS}
s(n) \sim n^{- b} \;, \qquad r(n) \sim c n^{- 1} \;,
\end{equation}
where $b$ and $c>0$ are constants. Thus the asymptotic forms of the
hop rates for large $n$ are given by
\begin{equation}
u(n,m) \sim (1-c/n^2)^m (1+b/n) \quad \textrm{and} \quad v(n,m) \sim 1+c/n\;.  
\end{equation}
We note that when $c=0$ the two species hop independently; in this
case, the asymptotic hop rates of the $A$ particles reduce to those
considered in \cite{E00} for the single species zero-range process, 
where condensation was found above a critical density provided $b>2$.

With the choice (\ref{FORMS}), condition (\ref{C1}) is
satisfied if $b>2$ and conditions (\ref{C2}) and (\ref{C3}) are satisfied
if $b > 3$. Therefore there are three cases to consider: 
 
\begin{description}
\item[Case 1:] $b<2$.

In this case, none of the conditions (\ref{C1}) to (\ref{C3}) is
met. The particular choice of rates studied in \cite{EH03} corresponds
to this case. The dependences of $\rho_A$ and $\rho_B$ on $y$ for fixed values
of $z$ are shown in Fig. \ref{fig:dense}, where $0< z_1 < z_2 < 1$.
\begin{figure} 
\begin{center}
\includegraphics[scale=0.5]{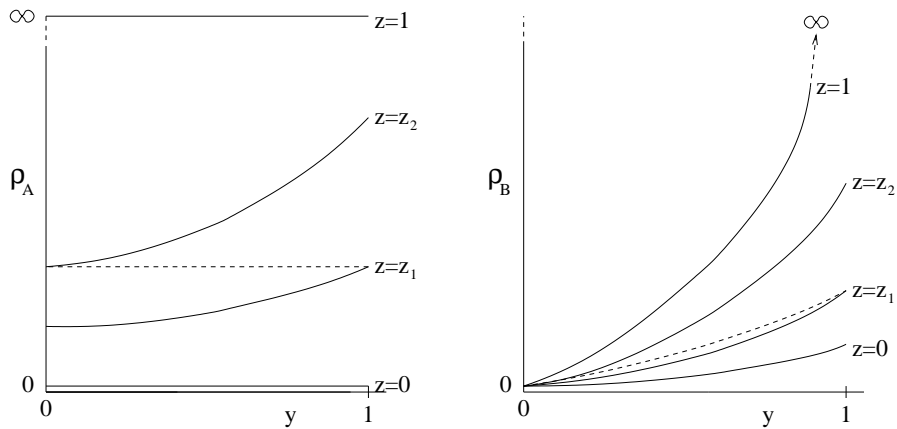}
\end{center} 
\caption{Schematic dependences for {\bf case 1}: $b<2$, of the particle densities $\rho_A$
and $\rho_B$ for 
contours of fixed $z$ and as a function of $y$. The dashed line
in the right hand graph
illustrates how $\rho_B$ varies as a function of $z$ and $y$ given
that $\rho_A$ is fixed (dashed line in left hand graph).} 
\label{fig:dense}
\end{figure}
Here, for a given $\rho_A$, $z$ must lie in the range $z_1
\leq z \leq z_2$. However, in this range, $\rho_B$ increases
monotonically from $\rho_B = 0$, where $y=0$ and $z=z_2$, to a maximum value at
$y=1$ and $z=z_1$. If $\rho_B$ exceeds this maximum then we can no
longer solve the saddle point equations (\ref{PD}) for both $\rho_A$
and $\rho_B$ and the excess $B$ particles condense onto a single
site. Therefore whenever $\rho_B$ exceeds a $\rho_A$-dependent
maximum, the system is in a condensate phase. Otherwise the system is
in a fluid phase. The critical line, given as a function of $z$ for
$y=1$, is shown in Fig. \ref{fig:pd1}. The explicit expression for
the critical line is
\begin{equation} \label{critline}
\rho_B = (1+\rho_A)/c \,.
\end{equation}
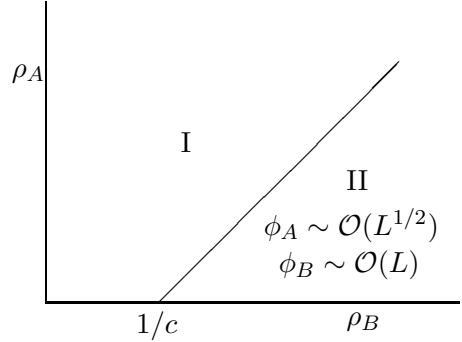
\begin{figure}
\begin{center}
\unitlength=1mm
\linethickness{0.4pt}
\begin{picture}(60,45)

\put(5,5){\line(1,0){55}}
\put(5,5){\line(0,1){40}}
\put(20,5){\line(1,1){32}}
\put(23,25){I}
\put(45,20){II}
\put(34,14){$\phi_A\sim\mathcal{O}(L^{1/2})$}
\put(36,9){$\phi_B\sim\mathcal{O}(L)$}
\put(0.5,35){$\rho_A$}
\put(45,2){$\rho_B$}
\put(17,1){$1/c$}

\end{picture}
\caption{Phase diagram for {\bf case 1}:
 $b<2$. Phase I is a fluid phase; in phase
II the $B$ particles form a condensate sustained by a `weak'
condensate of $A$ particles, as described in the text. $\phi_A$ and $\phi_B$
denote the numbers of particles contained in the condensates of $A$ and
$B$ particles respectively.} \label{fig:pd1}
\end{center}
\end{figure}
The condensate of $B$ particles (which contains $\mathcal{O}(L)$
particles) is induced by the distribution of $A$ particles. In
particular, at the site containing the $B$ particle condensate, the
$A$ particles form a `weak' condensate (which contains 
$\mathcal{O}(L^{1/2})$ particles). To see this, note
that the current of $A$ particles must be finite, therefore $u(n,m)$
must be finite at the condensate site. With the rates inferred from
(\ref{FORMS}), if $m\to \infty$ then $u(n,m)\to 0$ unless we also have
$n \to \infty$. Therefore taking $n$ large one finds that $u(n,m) \sim
{\rm exp}(-m/n^2)$. Since this must be finite we must
have $m\sim n^2$ at the condensate site. Then, because $m
\sim \mathcal{O}(L)$, we must have  $n\sim\mathcal{O}(L^{1/2})$. Away from the
condensate site, the $B$ particles form a power law distributed
background and the $A$ particles form an exponentially distributed
background. 

We can understand the weak condensate of $A$ particles by
considering a zero range process with a single defect site.
Consider a single species of particles --- the $A$ particles --- which
hop with rate $u(n)=1$ except at the defect site, where they hop with
rate $u(n)={\rm exp}(-g L/n^2)$ ($g$ is a constant). The hop rate from
the defect site reflects the effect of the $B$ particle condensate on
the $A$ particles in the two species
model. At the defect site, $u(n) \to 0$ if $n$ is small and so a
condensate forms. But since the hop rate must remain finite, the
condensate contains $n \sim\mathcal{O}(L^{1/2})$ particles. Hence,
from the perspective of the $A$ particles, the condensate of $B$
particles in the two species model plays the role of a defect site.

\item[Case 2:] $2<b<3$.

In this case, only condition (\ref{C1}) is met: $\rho_A$ is finite for
$z=1$ provided $y<1$. In this case, the dependences of $\rho_A$ and
$\rho_B$ on $y$ for fixed values of $z$ are shown in Fig.
\ref{fig:dense2}, again where $0< z_1 < z_2 < 1$.
\begin{figure} 
\begin{center}
\includegraphics[scale=0.5]{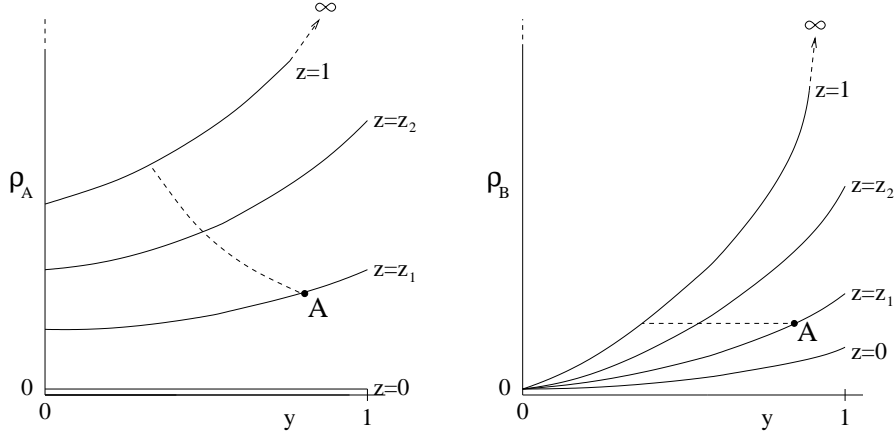}
\end{center} 
\caption{Schematic dependences for {\bf case 2}: $2<b<3$, of the particle densities $\rho_A$
and $\rho_B$ for 
contours of fixed $z$ and as a function of $y$. See text for commentary.} 
\label{fig:dense2}
\end{figure}
This time, imagine that the system is in the fluid phase, with
densities $\rho_A$ and $\rho_B$ (and therefore values of $z$ and $y$) 
corresponding to point $A$ in Fig.
\ref{fig:dense2}. Now, if we add more $A$ particles to the system
while keeping the density of $B$ particles fixed, we find that we must
increase $z$ and decrease $y$ as indicated by the dashed
line. However, when $z$ reaches 1, if we add more $A$ particles to the
system we can no longer solve the saddle point equations for $z$
and $y$, therefore the $A$ particles must undergo a transition from a
fluid phase to a condensate phase.
The critical line is given by $z=1$ --- the critical density of $A$
particles increases with 
increasing $\rho_B$. We have not been able to find an explicit
expression for this critical line. Also note that at $y=1$, the system
must undergo a transition between a fluid phase and a condensate of
$B$ particles as described in the previous case. Thus we deduce the
phase diagram shown in Fig. \ref{fig:pd2}. The two critical curves
intersect when $\rho_A=\infty$ and $\rho_B=\infty$. In phase III, the
condensate of $A$ particles 
exists on a power law distributed background of $A$ particles while
the $B$ particles are exponentially distributed throughout the system.
\begin{figure}
\begin{center}
\unitlength=1mm
\linethickness{0.4pt}
\begin{picture}(60,47)

\put(5,5){\line(1,0){55}}
\put(5,5){\line(0,1){42}}
\put(20,5){\line(1,1){32}}
\qbezier(5,20)(28,37)(49,42)
\put(20,20){I}
\put(45,20){II}
\put(34,14){$\phi_A\sim\mathcal{O}(L^{1/2})$}
\put(36,9){$\phi_B\sim\mathcal{O}(L)$}
\put(17,42){III}
\put(10,37){$\phi_A\sim\mathcal{O}(L)$}
\put(0.5,35){$\rho_A$}
\put(45,2){$\rho_B$}
\put(17,1){$1/c$}

\end{picture}
\caption{Phase diagram for {\bf case 2}: $2<b<3$. Phase I is a fluid phase; in phase
II the $B$ particles form a condensate sustained by a `weak'
condensate of $A$ particles, as described in the text. A condensate of
$A$ particles and fluid of $B$ particles form in phase III. $\phi_A$ and $\phi_B$
denote the numbers of particles contained in the condensates of $A$ and
$B$ particles respectively.} 
\label{fig:pd2}
\end{center}
\end{figure}
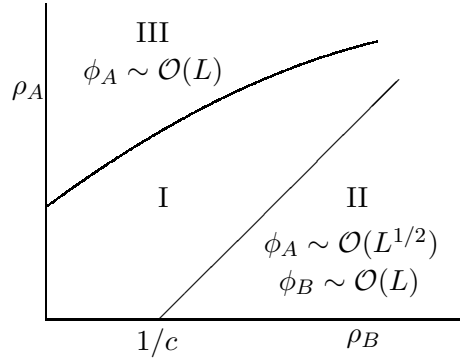
It is interesting to consider the sequence of transitions induced by
increasing $\rho_B$: starting from a point in phase III, the
condensate of $A$ particles is destroyed by increasing the
density of $B$ particles sufficiently, when the system enters the
fluid phase I. Increasing $\rho_B$ further leads the system to 
phase II where the $B$ particles condense.

\item[Case 3:] $b>3$

Here, all the conditions (\ref{C1}), (\ref{C2}) and (\ref{C3}) are
satisfied. The arguments of the previous two cases apply but now, the
critical curves given by $z=1$ on the one hand and 
$y=1$ on the other intersect at finite values of both $\rho_A$ and
$\rho_B$. Therefore when the $A$ and $B$ particle densities exceed
their values given by $z=1$ and $y=1$ the system enters a phase where
both species form a condensate at the same site. In this phase, the background
distributions of $A$ and $B$ particles are both given by power
laws. The phase diagram for this case is shown in Fig.
\ref{fig:pd3}.  
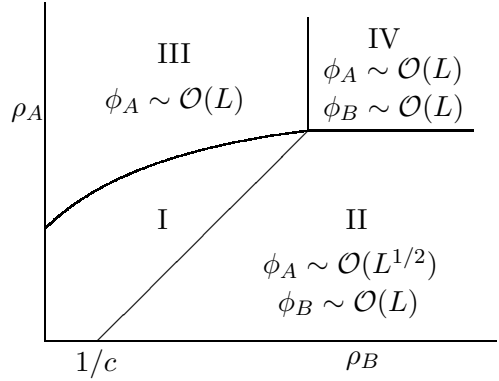
\begin{figure}
\begin{center}
\unitlength=1mm
\linethickness{0.4pt}
\begin{picture}(65,50)

\put(5,5){\line(1,0){60}}
\put(5,5){\line(0,1){45}}
\put(40,33){\line(1,0){22}}
\put(40,33){\line(0,1){15}}
\put(12,5){\line(1,1){28}}
\qbezier(5,20)(15,30)(40,33)
\put(20,20){I}
\put(45,20){II}
\put(34,14){$\phi_A\sim\mathcal{O}(L^{1/2})$}
\put(36,9){$\phi_B\sim\mathcal{O}(L)$}
\put(20,42){III}
\put(13,36){$\phi_A\sim\mathcal{O}(L)$}
\put(48,44){IV}
\put(42,40){$\phi_A\sim\mathcal{O}(L)$}
\put(42,35){$\phi_B\sim\mathcal{O}(L)$}
\put(0.5,35){$\rho_A$}
\put(45,2){$\rho_B$}
\put(9,1){$1/c$}

\end{picture}
\caption{Phase diagrams for {\bf case 3}: $b>3$. In region
I both species are in a fluid phase; in region II the $B$ particles
are in a condensate phase sustained by a weak condensate of $A$
particles; in region III the $A$ particles are in a condensate phase
and the $B$ particles form a fluid; in region IV both species are in a
condensate phase.
$\phi_A$ and $\phi_B$
denote the numbers of particles contained in the condensates of $A$ and
$B$ particles respectively.} \label{fig:pd3}
\end{center}
\end{figure}
 
\end{description}

\begin{figure}
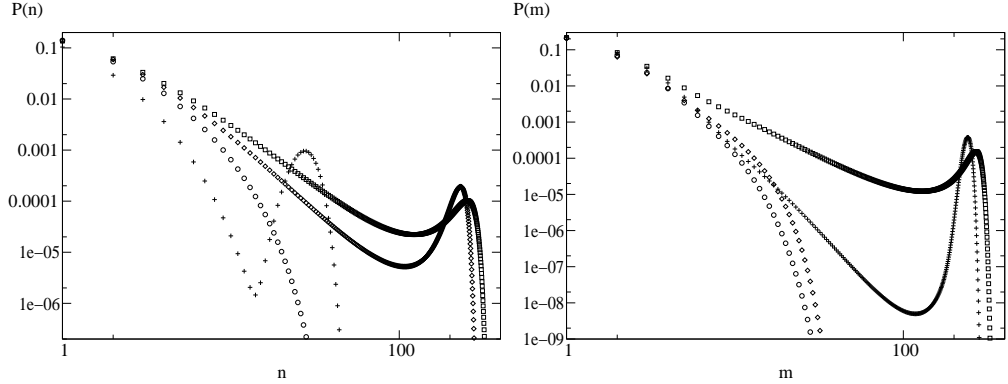
 
\begin{center}
\includegraphics[scale=0.27,angle=270]{PN.eps}
\includegraphics[scale=0.27,angle=270]{PM.eps}
\end{center} 
\caption{Log-log plot of on the left $P(n)$ vs. $n$, and on the right
$P(m)$ vs. $m$, for systems of size $L=100$ and $b=4$ and $c=2$. The circles
correspond to densities $\rho_A = 1/2 = \rho_B$ (fluid phase); the crosses
correspond to densities $\rho_A = 1/2$ and $\rho_B = 3$
(condensate of $B$ particles is sustained by a `weak' condensate of $A$
particles); the squares correspond to densities $\rho_A = 7/2 =
\rho_B$ (both species condense); the diamonds
correspond to densities $\rho_A = 3$ and $\rho_B = 1/2$ ($A$ particles
form a condensate).} 
\label{fig:dists}
\end{figure}

We have confirmed the existence
of the four phases presented in this section
numerically. Exact expressions for $P(n)$, the probability of finding
exactly $n$ $A$ particles at a site, and $P(m)$, the probability of finding
exactly $m$ $B$ particles at a site, can be obtained in terms of the
normalisation $Z_{L,N,M}$. This normalisation satisfies an exact
recursion equation \cite{EH03}
\begin{equation}
Z_{L,N,M} = \sum_{n=0}^N \sum_{m=0}^M f(n,m) Z_{L-1,N-n,M-m}\;,
\end{equation}
which is easily obtained from (\ref{ZLNM}), and which can be iterated on a
computer. Doing so, for systems up to size $L=100$ with $b=4$ and
$c=2$, yields the 
distributions shown in Fig. \ref{fig:dists}. Four phases are evident.
The circles represent densities $\rho_A = 1/2 = \rho_B$ and both
species are in a fluid phase (phase I). The crosses represent densities
$\rho_A = 1/2$ and $\rho_B = 3$, and the distributions are consistent
with a condensate of $B$ particles on a power law distributed
background, and a `weak' condensate of $A$ particles containing $n
\sim \mathcal{O} (L^{1/2})$ particles, on an exponentially distributed
background (phase II). The diamonds represent
densities $\rho_A = 3$ and $\rho_B = 1/2$  and the $A$ particles form
a condensate on a power law distributed background, while the $B$
particles form a fluid (phase III). The squares represent densities
$\rho_A = 7/2$ and $\rho_B = 7/2$ and both species form condensates on
power law distributed backgrounds (phase IV). 
  
It is possible to generalise the choice for $r(n)$ in (\ref{FORMS})
to, for example, $r(n) \sim c n^{-d}$ where $d>0$ is a constant. This
does not lead to any phase diagrams topologically distinct from those
already presented, although it does lead to the possibility of
$\rho_B$ converging to a finite value as $\rho_A \to \infty$, and vice
versa, as $z$ and $y \to 1$. Thus the phase diagrams in Figs.
\ref{fig:pd1} and \ref{fig:pd2} may be modified such that the phase
boundaries tend toward a finite value of $\rho_B$ as $\rho_A \to
\infty$, and vice versa also in the case of Fig. \ref{fig:pd2}. Another
feature of this generalised 
choice for $r(n)$ is that in the phase II, where the $B$ particles
condense, the accompanying `weak' condensate of $A$ particles contains a
number of particles $ n \sim \mathcal{O} (L^{1/(1+d)})$,
as may be verified using the argument expressed in case 1.


\section{Relation to AHR model}
\label{AHR}

In this section, we show that steady state of the two species
zero-range process has a mapping on to the steady state of the
AHR model.

The AHR model, introduced in \cite{AHR}, is a generalisation of
the second-class particle system studied in \cite{DJLS}.
It is defined on a ring of $L+N+M$ sites, on which there are $N$
$+$ particles, $M$ $-$ particles, and $L$ vacancies (which we represent
by $0$'s). The dynamics are defined by the processes 
\begin{eqnarray}
+\,0 &\rightarrow& 0\,+ \;, \qquad \textrm{with rate $\beta$}\;, \nonumber \\
0\,- &\rightarrow& -\,0 \;, \qquad \textrm{with rate $\alpha$} \;, \nonumber \\
+\,- &\rightarrow& -\,+ \;, \qquad \textrm{with rate 1}\;, \nonumber \\
-\,+ &\rightarrow& +\,- \;, \qquad \textrm{with rate $q$}\;,
\end{eqnarray}
where each exchange takes place between nearest neighbour sites. For
$\alpha = \beta = 1$ the model undergoes a transition between a
disordered phase ($q<1$) and a phase separated phase ($q>1$) composed
of a single domain of each species. The
correspondence between the AHR model and the two species zero-range
process may be observed in the following way.

We define $w(\{ \tau_i \})$ to be the steady state weight for the
system to be in a configuration $\{ \tau_i \} = \tau_1, \ldots,
\tau_{L+N+M}$. 
The weights $w(\{ \tau_i \})$ can be obtained using a matrix ansatz
\cite{AHR,DEHP,DJLS}, 
that is, we write the particle configuration as a product of matrices
\begin{equation} 
\{ \tau_i \} = X_1 \cdots X_{L+N+M}\;,
\end{equation}
where the matrix $X_i$ is
\begin{displaymath}
X_i = \left\{ 
\begin{array}{ll}
D & \textrm{if $\tau_i = +$}\;, \\
E & \textrm{if $\tau_i = -$}\;, \\
A & \textrm{if $\tau_i = 0$}\;.
\end{array} \right.
\end{displaymath}
Then it can be shown that the steady state weights can be written in
the form \cite{DEHP,DJLS} 
\begin{equation} \label{SSAHR}
w(\{ \tau_i \}) = \rm{Tr} [X_1 \cdots X_{L+N+M}]\;,
\end{equation}
provided the matrices $D$, $E$ and $A$ satisfy the relations
\begin{eqnarray}
\beta DA &=& A \;, \\
\alpha AE &=& A \;, \\
DE - qED &=& D+E \;.
\end{eqnarray}
These relations are satisfied if we take $A$ to be the projector
$\vert V \rangle\langle W \vert$, where we employ a bra-ket notation
to denote the left and right vectors $\langle W \vert$ and $\vert V
\rangle$. With this notation, (\ref{SSAHR}) becomes
\begin{equation}
w(\{ \tau_i \}) = \langle W \vert X_1 \cdots X_{L+N+M} \vert V \rangle\;,
\end{equation}
and if the $l$-th vacancy is at site $k_l$, then this can be written
(choosing the normalisation $\langle W \vert V \rangle = 1$ and using
the invariance of the trace under cyclic permutations of the $X$'s)
\begin{equation}
w(\{ \tau_i \}) = \prod_{l=1}^L \langle W \vert X_{k_l+1} \cdots
X_{k_{l+1}-1} \vert V \rangle\;, 
\end{equation}
i.e. the steady state weights assume a factorised form --- one factor
for each vacancy.
Now, to make the connection to the two species zero range process, we
define the matrix $G_{n,m}$ to be the sum over all permutations of products of 
$n$ $D$'s and $m$ $E$'s. Also, we define $P(\{n_l\};\{m_l\})$ to be the
probability that, in between all pairs of vacancies $l$ and $l+1$, there
are exactly $n_l$ $+$ particles and $m_l$ $-$ particles. Hence
\begin{equation} \label{AHRP}
P(\{n_l\};\{m_l\}) =  Z_{L,N,M}^{-1} \prod_{l=1}^L \langle W \vert
G_{n_l,m_l} \vert V \rangle\;, 
\end{equation}
where $Z_{L,N,M}$ is a normalisation. (\ref{AHRP}) is
identical to (\ref{PNM}) if we make the identification 
\begin{equation} \label{MAP}
f(n,m) = \langle W \vert G_{n,m} \vert V \rangle\;,
\end{equation}
and the normalisation $Z_{L,N,M}$ then is given by (\ref{ZLNM}). This
establishes the mapping.

Thus the steady state of the AHR model can be expressed in a form
identical to the steady state of the two species zero range process if
we identify the $+$ particles with the $A$ particles and the $-$
particles with the $B$ particles. The hop rates of the $A$ and
$B$ particles, obtained by substituting (\ref{MAP}) into (\ref{AB}),
are given by
\begin{equation}
u(n,m) = \frac{\langle W \vert G_{n-1,m} \vert V \rangle}{\langle W
\vert G_{n,m} \vert V \rangle}  \quad \textrm{and} \quad v(n,m) =
\frac{\langle W \vert G_{n,m-1} \vert V \rangle}{\langle W
\vert G_{n,m} \vert V \rangle}\;.
\end{equation}  
Note that because the mapping specifies $f(n,m)$ (and not the hop
rates) the hop rates are guaranteed to satisfy the constraint
equation (\ref{CE}). Also, this is not a mapping for the
microscopic dynamics ---
rather, it is a mapping between steady states which have the same
form.

The matrix elements $\langle W \vert G_{n,m} \vert V \rangle$ are
known exactly \cite{S} and assume different asymptotic forms depending
on the values of $\alpha$, $\beta$, and $q$. Thus the phase behaviour
of the AHR model is observed in the two species zero-range process by
using these different forms to determine $f(n,m)$ using
(\ref{MAP}). We note that the 
resulting values of the hop rates are different to those studied
earlier in this paper: the hop rates as determined via the mapping
to the AHR model obey the symmetry $u(n,m)=v(m,n)$ under the
interchange $\alpha \leftrightarrow \beta$. 


\section{Conclusion}
\label{C}

We have shown how the steady state of the two species zero-range
process can undergo a number of condensation transitions. A single
particle of one species was found to be able to induce condensation in
the other above a critical density. Next, for finite densities
of both species,  we investigated a case where the hop rates of the
two species were coupled in a nontrivial way.
Three distinct condensate phases emerged and the
conditions on the hop rates leading to such phases were presented for
quite general rates. This generality suggests that the
transition mechanisms are robust.

There remain a number of outstanding questions. A more detailed
understanding of the phase where the condensate is sustained by a `weak'
condensate of particles of the other species is desirable. It is also
unclear whether there exist further couplings between the particle
species which might lead to new transitions. This could require analysis of
the model for dynamics which do not satisfy the constraint (\ref{CE});
such investigation may also yield insight into the structure of the
steady state when the factorised form does not hold.

\vspace{5mm}
\noindent {\bf Acknowledgements}
We thank D. Mukamel for helpful discussions. TH thanks EPSRC for
financial support under grant GR/52497.


\begin{thebibliography}{00}

\bibitem{MKB}
S.N.~Majumdar, S.~Krishnamurthy and M.~Barma, Phys. Rev. Lett. {\bf
81}, 3691 (1998) 

\bibitem{KR}
P.L.~Krapivsky and S.~Redner, Phys. Rev. E {\bf 54}, 3553 (1996) 

\bibitem{E96}
M.R.~Evans, Europhys. Lett. {\bf 36}, 13 (1996) 

\bibitem{OEC}
O.J.~O'Loan, M.R.~Evans and M.E.~Cates, Phys. Rev. E {\bf 58}, 1404 (1998) 

\bibitem{AHR}
P.F.~Arndt, T. Heinzel and V.~Rittenberg, J.~Phys.~A {\bf 31}, L45
(1998); J.~Stat.~Phys. {\bf 97}, 1 (1999) 

\bibitem{M}
D.~Mukamel, in {\it Soft and Fragile Matter: Nonequilibrium Dynamics,
Metastability and Flow}, edited by M.E.~Cates and M.R.~Evans (Institute
of Physics Publishing, Bristol, 2000)

\bibitem{KLMST}
Y.~Kafri, E.~Levine, D.~Mukamel, G.M.~Sch{\"u}tz and J.~T{\"o}r{\"o}k,
Phys.~Rev.~Lett. {\bf 89}, 035702 (2002) 

\bibitem{E00}
M.R.~Evans, Braz. J. Phys. {\bf 30}, 42 (2000)  

\bibitem{EH03}
M.~R.~Evans and T.~Hanney, J.~Phys.~A {\bf 36}, L441 (2003) 

\bibitem{GS03}
S.~Gro\ss kinsky and H.~Spohn, {\it Preprint} cond-mat/0305306
   
\bibitem{LR}
R.~Lahiri and S.~Ramaswamy, Phys.~Rev.~Lett. {\bf 79}, 1150 (1997) 
 
\bibitem{DK}
B.~Drossel and M.~Kardar, Phys.~Rev.~Lett. {\bf 85}, 614 (2000) 

\bibitem{DB}
D.~Das and M.~Barma, Phys.~Rev.~Lett. {\bf 85}, 1602 (2000) 

\bibitem{DEHP}
B.~Derrida, M.R.~Evans, V.~Hakim and V.~Pasquier, J.~Phys.~A {\bf 26},
1493 (1993) 

\bibitem{DJLS}
B.~Derrida, S.A.~Janowsky, J.~L.~Lebowitz and E.R.~Speer,
J.~Stat.~Phys. {\bf 73}, 813 (1993) 
 
\bibitem{S}
K.~Mallick, J. Phys. A {\bf 29}, 5375 (1996);
T.~Sasamoto, Phys. Rev. E {\bf 61}, 4980 (2000) 

\end{thebibliography}
\end{document}